# Two-dome structure in electron-doped iron arsenide superconductors

Soshi Iimura,[1] Satoru Matuishi,[1] Hikaru Sato,[1] Taku Hanna,[1] Yoshinori Muraba,[1] Sung Wng Kim,[2,**] Jung Eun Kim,[3] Masaki Takata,[3,4] and Hideo Hosono[1,2,*]

[1]*Materials and Structures Laboratory, Tokyo Institute of Technology, 4259 Nagatsuta-cho, Midori-ku, Yokohama 226-8503, Japan*

[2]*Frontier Research Center, Tokyo Institute of Technology, 4259 Nagatsuta-cho, Midori-ku, Yokohama 226-8503, Japan*

[3]*Japan Synchrotron Radiation Research Institute, 1-1-1 Kouto, Sayo-cho, Sayo-gun, Hyogo 679-5198, Japan*

[4]*RIKEN SPring-8 Center, 1-1-1 Kouto, Sayo-cho, Sayo-gun, Hyogo 679-5148, Japan*

e-mail: hosono@lucid.msl.titech.ac.jp

**Present address: Department of Energy Science, Sungkyunkwan University, 300 Cheoncheon, Jangan-ku, Suwon, Korea

**Iron arsenide superconductors based on the material LaFeAsO$_{1-x}$F$_x$ are characterized by a two-dimensional Fermi surface (FS) consisting of hole and electron pockets yielding structural and antiferromagnetic transitions at $x = 0$. Electron doping by substituting O$^{2-}$ with F$^-$ suppresses these transitions and gives rise to superconductivity with a maximum $T_c$ = 26 K at $x = 0.1$. However, the over-doped region cannot be accessed due to the poor solubility of F$^-$ above $x = 0.2$. Here we overcome this problem by doping LaFeAsO with hydrogen. We report the phase diagram of LaFeAsO$_{1-x}$H$_x$ ($x < 0.53$) and, in addition to the conventional superconducting dome seen in LaFeAsO$_{1-x}$F$_x$, we find a second dome in the range $0.21 < x < 0.53$, with a maximum $T_c$ of 36 K at $x = 0.3$. Density functional theory calculations reveal that the three Fe 3$d$ bands ($xy$, $yz$, $zx$) become degenerate at $x = 0.36$, whereas the FS nesting is weakened monotonically with $x$. These results imply that the band degeneracy has an important role to induce high $T_c$.**

Since the discovery of superconductivity in LaFeAsO1 − xFx with $T_c$ = 26 K in early 2008 (ref. 1), various types of iron pnictides containing square lattices of Fe2 + have been investigated[2–4]. That maximum was raised to 55 K in *Ln*-1111-type *Ln*FeAsO1 − xFx (*Ln* denotes lanthanide)[5]. The compound of LaFeAsO1 − xFx is paramagnetic metal with tetragonal symmetry at room temperature and undergoes a tetragonal–orthorhombic transition around 150 K accompanied by a para-antiferromagnetic (AFM) transition[6,7]. Superconductivity emerges when the transitions are suppressed by carrier doping via element substitution or pressure application. To explain the emergence of superconductivity near the AFM phase, a spin fluctuation model resulting from Fermi surface (FS) nesting between hole and electron pockets was proposed based on density functional theory (DFT) calculations[8,9]. This model explains the suppression of superconductivity in LaFeAsO1 − xFx upon electron doping to the filling level of hole pockets ($x$ = 0.2) and the striking difference in the maximum $T_c$ between *Ln*FeAsO1 − xFx (26–55 K) and LaFePO (4 K) or between La-1111(26 K) and Sm-1111(55 K)[10].

However, the phase diagrams reported so far for *Ln*FeAsO systems are rather incomplete. For instance, the suppression of $T_c$ in over-doping region had not been confirmed for any *Ln*-1111-types (except La). This situation primarily comes from the low solubility limit of fluorine in $Ln$FeAsO1 − $x$F$x$ ($x$ < 0.15–0.20). Recently, we reported the syntheses of (Ce, Sm)FeAsO1 − $x$H$x$ (0 < $x$ < 0.5) by using the high solubility limit of hydrogen and obtained a complete superconducting dome ranging 0.05 < $x$ ≤0.4~0.5 with optimum $T$c of 47 K for the Ce-system or 56 K for the Sm-system, agreeing well with that the previous data of each fluorine-doped sample in $x$ < 0.15 (refs 11,12). The position and occupancy of hydrogen substituting the oxygen sites were verified by neutron powder diffraction measurement and the charge state of hydrogen was examined by DFT calculations12. The neutron powder diffraction measurement on CeFeAsO1 − $x$D$x$ revealed that hydrogen species exclusively substitute the oxygen sites in the CeO layer, and DFT calculations indicated that the hydrogen 1$s$ band is located at − 3 to − 6 eV, which is close to level of the oxygen 2$p$ band and the charge state of the hydrogen is − 1. These results substantiate the idea that hydrogen exclusively substituting the O2– sites occurs as H − , supplying electrons to the FeAs layer, the same as fluorine does (O2– = H − + e − ).

When comparing the superconducting dome of LaFeAsO1 − $x$F$x$ with (Ce, Sm)FeAsO1 − $x$H$x$, it is found that the dome width is twice as narrow as that of (Ce, Sm)FeAsO1 − $x$H$x$ and the optimum $T$c of LaFeAsO1 − $x$F$x$ is much lower than that of (Ce, Sm)FeAsO1 − $x$H$x$. Moreover, the temperature dependence of resistivity in the normal conducting state indicates that the (Ce, Sm)FeAsO$_{1-x}$H$_x$ behave non-Fermi liquid, that is, $\rho(T) \sim T$, whereas LaFeAsO$_{1-x}$F$_x$ obeys Fermi liquid, that is, $\rho(T) \sim T^2$. These differences remind us the idea that the electron doping via fluorine substitution in LaFeAsO$_{1-x}$F$_x$ is not enough to draw out the genuine physical property of LaFeAsO. In this Article, we study the LaFeAsO system, the prototype material for the *Ln*-1111 system, and examine its superconducting properties by using H − in place of F − for electron doping. We report the existence over a wider *x* range of another *T*c dome with higher *T*c in addition to the dome reported so far by F − substitution. Not only are there differences in maximum *T*c and shape between these two domes, but also the second dome with resistivity characterized by a linear temperature dependence corresponds to that observed in other *Ln*-1111 systems (except La) with higher *T*c. Based on DFT calculations of the crystal structures of these doped samples determined at 20 K, we discuss the origin for the superconductivity.

**Results**

**Two-dome structure**. In Fig. 1a,b, we show the temperature dependence of electrical resistivity for LaFeAsO1 − *x*H*x*. At *x* = 0.01 and 0.04, a kink in resistivity due to structural or magnetic transitions was seen around 150 K (refs 6,7). As *x* is increased, the transitions are suppressed and the onset *T*c appears for *x* ≥ 0.04, and zero resistivity is attained for *x* ≥ 0.08. The onset *T*c, determined from the intersection of the two extrapolated lines in Fig. 1c,d attains a maximum of 29 K at *x* = 0.08 and decreases to 18 K at *x* = 0.21, forming the first *T*c (*x*) dome. For 0.08≤ *x* ≤0.21, the temperature dependence of resistivity ρ(*T*) in the normal state obeys a *T*2-law, indicating that the system is a strongly correlated metal in Fermi liquid theory13. The *T*c (*x*) and temperature dependence of ρ above *T*c agree well with those in LaFeAsO1 − *x*F*x* (ref. 1). Further electron doping (*x* > 0.21) continuously enhances *T*c to 36 K around *x* = 0.36 for which ρ(*T*) exhibits *T*-linear dependence. Figure 1e,f shows the temperature dependence of volume magnetic susceptibility χ for samples with different *x*. Because of the presence of metal iron impurities, each sample has a positive offset. The diamagnetism due to superconductivity is clear to see for *x* >0.04. Shielding volume fraction exceeds 40% at 2 K in 0.08≤ *x* ≤0.46, and then decreases to 20% at *x* = 0.53, forming the second $T$c dome in the composition range $0.21 < x < 0.53$. This dome is newly found by the present study.

**Pressure effect**. Figure 2a–d shows the changes in $\rho(T)$ for samples with $x = 0.08$, 0.21, 0.30 and 0.46 under applied pressures ($P$) up to 3 GPa. The onset $T$c increases largely with $P$ for $x = 0.08$, 0.21 and 0.30, whereas the $T$c for $x = 0.46$ slightly decreases from 33 K to 32 K at $P = 2.7$ GPa. The maximum $T$c obtained at $x = 0.30$ under $P = 2.6$ GPa is 46 K, which is distinctly higher than the maximum $T$c (43 K) in the LaFeAsO1 − $x$F$x$ under high pressure14. The $T$c valley around $x = 0.21$ under ambient pressure disappears, resulting in the one-dome structure as observed in other $Ln$-1111 series. Figure 2e summarizes the $T$s and $T$c under ambient pressure in the LaFeAsO1 − $x$H$x$ and LaFeAsO1 − $x$F$x$ (ref. 15) along with $T$c at $P = 3$ GPa of LaFeAsO1 − $x$H$x$. Two superconducting domes are evident and each has maximum $T$c around $x = 0.08$ and 0.36. The first dome is located adjacent to the orthorhombic and AFM phase and is almost the same as previously reported for LaFeAsO1 − $x$F$x$, whereas the second dome appears adjacent to the first dome. At $P = 3$ GPa, the two domes merge into a wider dome having a closed shape and a range similar to those in CeFeAsO1 − $x$H$x$ with maximum $T$c = 47 K (ref. 12). This unification of the two domes on applying high pressure may be understood as lattice compression; the reduction of the $a$-axis (~1% under 3 GPa) in LaFeAsO1 − $x$F$x$ is assumed to bring it close to a LaFeAsO1 − $x$H$x$ lattice16. Then, a 1%-reduction of *a*-axis for LaFeAsO1 − $x$H$x$ draws its lattice close to that of CeFeAsO1 − $x$H$x$ under an ambient pressure12.

**Lanthanide cation substitution effect**. Here we consider the relation between the two domes of the La-system and the domes of the other *Ln*-1111 systems. To compare the temperature dependence of resistivity of LaFeAsO1 − $x$H$x$ with that of other *Ln*-1111 systems, we performed power-law fitting, ρ = ρ0 + $ATn$ (ρ0: residual resistivity) in the temperature range above *T*c to 150 K (Supplementary Fig. S1). Figure 3a shows the relation between the exponents *n* and *x* for *Ln*FeAsO1 − $x$H$x$ (*Ln* = La, Ce, Sm and Gd; The sample preparation and temperature dependence of electrical resistivity and volume magnetic susceptibility for newly found GdFeAsO1 − $x$H$x$ are summarized in Supplementary Fig. S2 and in the Supplementary Methods.). Fermi liquid-like behaviour, *n*=2, is observed only in low-*x* LaFeAsO1 − $x$H$x$, whereas non-Fermi liquid behaviuor, *n* < 2, is observed for high-*x* LaFeAsO1 − $x$H$x$ and for the entire range of *x* in the other systems. Figure 3b shows the plot of *T*c versus exponent *n* for the same systems. As *n* approaches unit for each system, *T*c becomes a maximum, indicating that this feature of the second dome in LaFeAsO1 − $x$H$x$ is commonly seen for domes in other *Ln*-1111; that is, the first dome is unique to La-1111, whereas the second is universally to all four systems. Figure 3c–f shows the *x*-variation in $T_c$ for all four systems. The optimal *x* in the $T_c$ dome continuously shifts to lower *x* when comparing *Ln* = La through to Gd.

**Discussion**

We have found an unusual two-dome structure in $T_c$ for the LaFeAsO1 − *x*H*x* system, the higher $T_c$ dome being associated with a universal structure of *Ln*FeAsO1 − *x*H*x* systems generally. To understand these dome structures, we calculated the electronic state of these materials using the crystal structures determined at 20 K. The electron doping via substitution of O2– sites with H − ion was modelled in virtual crystal approximation assuming hydrogen acts as a quasi-fluorine ion supplying an electron to the FeAs layer12, that is, the oxygen ($Z = 8$) sites were substituted for virtual atoms which have a fractional nuclear charge ($Z = 8 + x$), where *x* is hydrogen fraction. Figure 4a–d shows the two-dimensional cross-sections of FS for the various doping levels. These compositions, $x = 0.08, 0.21, 0.36$ and $0.40$, correspond to the top of first dome, $T_c$ valley, the top of the second dome and over-doping region, respectively. At $x = 0.08$, the size of an outer $d_{xy}$ (*x*, *y* and *z* coordination is given by the Fe square lattice) or inner $d_{yz,zx}$ hole pockets at the Γ point is close to that of two electron pockets at the M point, indicating that nesting in the (π π) direction between the hole and the electron pockets is strong. As *x* increases, the nesting monotonically weakens because the hole pockets are gradually reduced but the electron pockets are expanded. It is pointed out that as the pnictogen height, $h$Pn, from the Fe plane increases, the *dxy* hole pocket is enlarged; nesting then becomes better10. In the present case, although $h$As increases with *x* as shown in Fig. 4e, the size of the *dxy* hole pocket remains almost unchanged irrespective of *x*. This result may be understood by considering that expansion of *dxy* hole pocket by structural modification is cancelled by reduction due to the up-shift of Fermi level ($E_F$) by electron doping.

Nesting between hole and electron pockets is the most important glue in the spin fluctuation model8,9. The decrease in $T_c$ from $x = 0.08$ through 0.21 may be understood as a reduction in spin fluctuations due to weakening of the nesting in a similar manner to LaFeAsO$_{1-x}$F$_x$. It is, however, difficult to understand by the FS nesting that the experimental findings that $T_c$ (*x*) increases over a wider dome range of $0.21 < x < 0.53$ and optimizes at 36 K around $x = 0.36$.

Figure 4f–i shows band structures near $E_F$ for sample composition $x = 0.08$, 0.21, 0.36 and 0.40. As the unit cell contains two irons, there are ten bands around the $E_F$ derived from bonding and anti-bonding of 3*d* orbitals of the two neighbouring irons that cross the $E_F$ around the Γ and M points. The unoccupied bands move continuously

lower with *x*. In particular, the bands derived from the anti-bonding orbital between the *dxy* orbitals, which we shall call the 'anti-*dxy* bands' hereafter, and the band derived from the degenerate *dyz,zx* bonding orbitals are lowered in energy and cross the bonding-*dxy* band around $x = 0.36$, forming degenerate states of Fe 3*dxy,yz,zx* three bands near $E_F$ as seen in Fig. 4j. After this triply-degenerate state is formed, the anti-*dxy* band and bonding *dyz,zx* band create a new band below $E_F$ at $x = 0.40$ by reconstruction (see inset of Fig. 4i). Note in Fig. 4e that the As–Fe–As angle of FeAs4 tetrahedron is far from that of regular tetrahedron (109.5°) in the optimally doped region ($x = 0.33 - 0.46$). The band structures of LaFeAsO1 − *x*H*x* ($x = 0.08, 0.21, 0.36$ and $0.40$) with only structural change are shown in Supplementary Fig. S3. Although the magnitude of the energy difference between these three bands becomes small with *x*, the band crossing is not caused only by the structural change, indicating that not only the change in local structure around iron but also asymmetric occupation of doped electrons in the bonding-*dyz,zx*, *dxy* and anti-bonding-*dxy,* affect the band shift. As the bonding *dyz,zx* and *dxy* bands are almost flat along the Γ-Z direction, their band crossing at $x = 0.36$ form a shoulder in the total density of states (DOS) at $E_F$ (Fig. 4m), indicating an electronic instability of the system arising from degeneration of the *dxy* and *dyz,zx* bands. In such a situation, structural transitions, for instance band Jahn-Teller distortion, may occur to reduce the energy of the system. However, in the present results, any structural transition could not be observed at least down to 20 K in samples with $0.08 \leq x \leq 0.40$.

Table 1 summarizes the characteristics of each $T$c dome described above. The primary question is what the origin is for the second dome, that is, the dome in the *Ln*-1111 with higher *T*c. The resistivity above *T*c changes from quadratic to linear dependence as *x* approaches the top of the second dome region. As the nesting between hole and electron pockets monotonically is weakened with *x*, it is rationally considered that the contribution of FS nesting to the second *T*c dome is not dominant. In addition, the calculated DOS shows the presence of a shoulder of the DOS (*E*F) at $x = 0.36$ related by the degeneracy of three bands derived from Fe-*dyz*,zx and *dxy* orbitals. Given these results, the band degeneracy appears to have an important role in emergence of the second dome. For the iron-based superconductors, there is another pairing model derived from a large softening of a shear modulus observed near the tetragonal–orthorhombic transition of parent compounds[17–19]. This model tells that the Fe-*d* orbitals are possible to order when their degeneracy in Fe-*dyz*,zx orbitals is removed at the structural transition and the fluctuations of this orbital ordering are shown capable of inducing superconductivity[18,19]. If we follow the orbital fluctuation model, the second dome and *T*-linear resistivity in the present system might be understood as results of electron pairing and carrier scattering by the fluctuations of the degenerated Fe-*dxy,yz,zx* orbitals, respectively.

Finally, we consider why the two-dome structure is only found in LaFeAsO1 − *x*H*x* and not (Ce-Gd)FeAsO1 − *x*H*x*. The degeneracy of three bands derived from Fe-3*dyz,zx* and *dxy* orbitals is realized when energy splitting between the 3*dyz,zx* and *dxy* bands mainly derived from distortion of FeAs4 tetrahedron is cancelled out by asymmetric occupation of doped electrons in the these three bands. The magnitude of the energy difference between these three bands becomes small on going from La ion to Ce–Gd ions in *Ln*FeAsO because the As–Fe–As angle of FeAs4 tetrahedron continuously approach to the angle (109.5°) of a regular tetrahedron as the lanthanide ion is changed from La (114°) to Gd (110°)20. In particular, this deviation in the La-system is rather large compared with the other system. Therefore, the threefold degeneracy in (Ce-Gd)FeAsO1 − *x*H*x* on electron doping is expected to occur in lower *x* than that (*x* = 0.35) in LaFeAsO1 − *x*H*x*. As a consequence, we think that the second dome primarily originating from the band degeneracy is separated from the first dome from FS nesting. The present discussion on orbital fluctuation model is based on DFT calculations. Further effort is required to confirm the validity of this idea on decisive experimental evidences such as angular-resolved photoemission and elastic shear modulus measurements using the single crystals.

**Methods**

**Synthesis, structural and chemical analyses of LaFeAsO1 − $x$H$x$**. As reported previously, LaFeAsO1 − $x$H$x$ was synthesized by solid-state reactions using starting materials La2O3, LaAs, LaH2, FeAs and Fe2As under high pressure11. The phase purity and structural parameters at room temperature were determined by powder X-ray diffraction measurements with MoKα1 radiation. The structure parameters at low temperatures were determined by synchrotron X-ray diffraction measurements at 20 K using the BL02B2 beam-line in the SPring-8, Japan. Hydrogen content in the synthesized samples was evaluated by thermogravimetric mass spectroscopy, and the chemical composition with the exception of hydrogen was determined with a wavelength-dispersive-type electron-probe microanalyzer.

**Electric and magnetic measurement**. Resistivity and magnetic susceptibility at ambient pressure were measured using a physical property measurement system with a vibrating sample magnetometer attachment. Electrical resistivity measurements under high pressure were performed by the dc four-probe method. Pressures were applied at room temperature and maintained using a piston-cylinder device. A liquid pressure-transmitting medium (Daphne oil 7373) was used to maintain hydrostatic conditions.

**Density functional theory calculations**. The electronic structure for LaFeAsO$_{1-x}$H$_x$ was derived from non-spin-polarized DFT calculations using the WIEN2K code[22] using the generalized gradient approximation Perdew–Burke–Ernzerhof functional[23] and the full-potential linearized augmented plane wave plus localized orbitals method. To ensure convergence, the linearized augmented plane wave basis set was defined by the cutoff $R$MT$K$MAX = 9.0 ($R$MT: the smallest atomic sphere radius in the unit cell), with a mesh sampling of 15×15×9 $k$ points in the Brillouin zone.

**References**

1. Kamihara, Y., Watanabe, T., Hirano, M. & Hosono, H. Iron-based layered superconductor La[O$_{1-x}$F$_x$]FeAs ($x$=0.05–0.12) with $T_c$=26 K. *J. Am. Chem. Soc.* **130,** 3296–3297 (2008).

2. Rotter, M., Tegel, M. & Johrendt, D. Superconductivity at 38 K in the iron arsenide (Ba$_{1-x}$K$_x$)Fe$_2$As$_2$. *Phys. Rev. Lett.* **101,** 107006 (2008).

3. Wang, X. C. *et al.* The superconductivity at 18 K in LiFeAs system. *Solid State Commun.* **148,** 538–540 (2008).

4. Zhu, X. *et al.* Transition of stoichiometric Sr$_2$VO$_3$FeAs to a superconducting state at 37.2 K. *Phys. Rev. B* **79,** 220512 (2009).

5. Ren, Z. A. *et al.* Superconductivity at 55 K in iron-based F-doped layered quaternary compound Sm[O1 − *x*F*x*] FeAs. *Chin. Phys. Lett.* **25,** 2215–2216 (2008).

6. de la Cruz, C. *et al.* Magnetic order close to superconductivity in the iron-based layered LaO1 − *x*F*x*FeAs systems. *Nature* **453,** 899–902 (2008).

7. Nomura, T. *et al.* Crystallographic phase transition and high-*T*c superconductivity in LaFeAsO: F. *Supercond. Sci. Technol.* **21,** 125028 (2008).

8. Mazin, I. I., Singh, D. J., Johannes, M. D. & Du, M. H. Unconventional superconductivity with a sign reversal in the order parameter of LaFeAsO1 − *x*F*x*. *Phys. Rev. Lett.* **101,** 05700 (2008).

9. Kuroki, K. *et al.* Unconventional pairing originating from the disconnected fermi surfaces of superconducting LaFeAsO1 − *x*F*x*. *Phys. Rev. Lett.* **101,** 087004 (2008).

10. Kuroki, K., Usui, H., Onari, S., Arita, R. & Aoki, H. Pnictogen height as a possible switch between high-*T*c nodeless and low-*T*c nodal pairings in the iron-based superconductors. *Phys. Rev. B* **79,** 224511 (2009).

11. Hanna, T. *et al.* Hydrogen in layered iron arsenides: indirect electron doping to induce superconductivity. *Phys. Rev. B.* **84,** 024521 (2011).

12. Matsuishi, S. *et al.* Structural analysis and superconductivity of CeFeAsO1 − *x*H*x*. *Phys. Rev. B.* **85,** 014514 (2012).

13. Imada, M., Fujimori, A. & Tokura, Y. Metal-insulator transitions. *Rev. Mod. Phys.* **70,** 1039 (1998).

14. Takahashi, H. *et al.* Superconductivity at 43 K in an iron-based layered compound LaO1 − *x*F*x*FeAs. *Nature* **453,** 376–378 (2008).

15. Hess, C. *et al.* The intrinsic electronic phase diagram of iron-oxypnictide superconductors. *Europhys. Lett.* **87,** 17005 (2009).

16. Takahashi, H. *et al.* High-pressure studies on superconducting iron-based LaFeAsO1 − *x*F*x*, LaFePO and SrFe2As2. *J. Phys. Soc. Jpn.* **77,** 78 (2008).

17. Yoshizawa, M. *et al.* Structural quantum criticality and superconductivity in iron-based superconductor Ba(Fe1 − *x*Co*x*)2As2. *J. Phys. Soc. Jpn.* **81,** 024604 (2012).

18. Kontani, H. & Onari, S. Orbital-fluctuation-mediated superconductivity in iron pnictides: analysis of the five-orbital hubbard-holstein model. *Phys. Rev. Lett.* **104,** 157001 (2010).

19. Yanagi, Y., Yamakawa, Y., Adachi, N. & Ōno, Y. Orbital order, structural transition, and superconductivity in iron pnictides. *J. Phys. Soc. Jpn.* **79,** 123707 (2010).

20. Wang, P., Stadnik, Z. M., Wang, C., Cao, G. H. & Xu, Z. A. Transport, magnetic, and 57Fe and 155Gd Mössbauer spectroscopic properties of GdFeAsO and the slightly overdoped superconductor Gd0.84Th0.16FeAsO. *J. Phys.:Condens. Matt.* **22,** 145701

(2010).

21. Haverkort, M. W., Elfimov, I. S., Tjeng, L. H., Sawatzky, G. A. & Damascelli, A. Strong spin-orbit coupling effects on the fermi surface of Sr2RuO4 and Sr2RhO4. *Phys. Rev. Lett.* **101,** 026406 (2008).

22. Blaha, P. *et al. An Augmented Plane Wave and Local Orbitals Program for Calculating Crystal Properties*, Technical University of Wien, Vienna( 2001).

23. Perdew, J. P., Burke, K. & Ernzerhof, M. Generalized gradient approximation made simple. *Phys. Rev. Lett.* **77,** 3865 (1996); 78, 1396(E) (1997).

**Acknowledgements**

We thank Professor H. Fukuyama of Tokyo university of Science for discussions. This research was supported by the Japan Society for the Promotion of Science (JSPS) through the FIRST program, initiated by the CSTP. The synchrotron radiation experiments were performed at the BL02B2 of SPring-8 with the approval of the Japan Synchrotron Radiation Research Institute (JASRI; proposal no. 2011A1142).

**Author contributions**

H.H. and S.M. planned the research. S.I., T.H. and Y.M. performed the high-pressure synthesis. S.I. performed measurement. H.S. and S.W.K. carried out high-pressure resistivity measurement. S.I., J.E.K. and M.T. performed Synchrotron X-ray diffraction measurements. S.I. and S.M. performed DFT calculations. H.H. and S.I. and S.M. discussed the results and wrote the manuscript.

**Additional information**

**Supplementary Information** accompanies this paper at http://www.nature.com/naturecommunications

**Competing financial interests:** The authors declare no competing financial interests.

**Reprints and permission** information is available online at http://npg.nature.com/reprintsandpermissions/

**How to cite this article:** Iimura, S *et al.* Two-dome structure in electron-doped iron arsenide superconductors. *Nat. Commun.* 3:943 doi: 10.1038/ncomms1913 (2012).

**Figure caption**

**Figure 1 | Electrical and magnetic properties of LaFeAsO1 − *x*H*x*.** (**a**,**b**) Electrical resistivity as a function of temperature in $x$ = 0.01–0.21 (**a**) and 0.24–0.53 (**b**). (**c**,**d**) Enlarged ρ–*T* curves near *T*c of LaFeAsO1 − *x*H*x* with $x$ = 0.08–0.21 (**c**) and 0.24–0.53 (**d**). The arrows mark the onset *T*c. (**e**,**f**) Magnetic susceptibility data in $x$ = 0.01–0.21 (**e**) and 0.24–0.53 (**f**) measured with a zero-field-cooling history and a field of 10 Oe.

**Figure 2 | Temperature dependence of the electrical resistivity of LaFeAsO1 − *x*H*x* below 3 GPa and phase diagrams of LaFeAsO1 − *x*(H, F)*x*.** (**a**–**d**) Temperature dependence of resistivity as a function of static pressure at $x$ = 0.08 (**a**), 0.21 (**b**), 0.30 (**c**) and 0.46 (**d**). The insets show pressure dependence of *T*c. (**e**) Electronic phase diagram for LaFeAsO1 − *x*H*x* (filled symbols) and LaFeAsO1 − *x*F*x* (ref. 15; open symbols). The *T*c under ambient (squares) and 3 GPa (inverted triangles) was determined from the intersection of the two extrapolated lines around superconducting transition and *T*s (circles) was taken as the anomaly kink in the ρ–*T* curve7.

**Figure 2 | Temperature dependence of the electrical resistivity of LaFeAsO1 − xHx below 3 GPa and phase diagrams of LaFeAsO1 − x(H, F)x.** (**a**–**d**) Temperature dependence of resistivity as a function of static pressure at *x* = 0.08 (**a**), 0.21 (**b**), 0.30 (**c**) and 0.46 (**d**). The insets show pressure dependence of *T*c. (**e**) Electronic phase diagram for LaFeAsO1 − xHx (filled symbols) and LaFeAsO1 − xFx (ref. 15; open symbols). The *T*c under ambient (squares) and 3 GPa (inverted triangles) was determined from the intersection of the two extrapolated lines around superconducting transition and *T*s (circles) was taken as the anomaly kink in the ρ–*T* curve7.

**Figure 4 | Electronic structure of LaFeAsO1 − xHx.** (**a**–**d**) Two-dimensional Fermi surface of LaFeAsO1 − xHx with *x* = 0.08 (**a**), 0.21 (**b**), 0.36 (**c**) and 0.40 (**d**). The blue arrow represents the nesting vector in the (π π) direction. The contribution of Fe-*dxy* and *dyz,zx* orbitals are coloured green and pink, respectively21. (**e**) Hydrogen doping dependence of As–Fe–As angle in FeAs4 tetrahedron (orange circles) and arsenic height *h*As from the Fe plane (dark sky-blue diamonds). The angles and *h*As are determined from Synchrotron X-ray diffraction measurements at 20 K. The solid lines are as a visual guide. The dashed line denotes the regular tetrahedron angle. (**f**–**i**) Band structures of LaFeAsO1 − xHx with *x* = 0.08 (**f**), 0.21 (**g**), 0.36 (**h**) and 0.40 (**i**). Insets show close-up views of the low energy region. The contribution of Fe-*dxy* and *dyz,zx*

orbitals are coloured green and pink, respectively. (**j**) Variation in energy level of relevant Fe 3*d* bands at Γ point with *x*. The inset is the band structure of LaFeAsO$_{0.92}$H$_{0.08}$. The Γ*dxy* (filled green inverted triangles) and Γanti-*dxy* (open green triangles) signify the bonding and anti-bonding states, respectively, for a bond primary composed of two Fe *dxy* orbitals in a unit cell. Also shown is the energy level of degenerate *dyx,zx* band (Γ*dyz,zx* indicated by filled pink squares). The solid and dashed lines are as a visual guide. (**k**–**n**) Total DOS(solid black line) and partial DOSof *dxy* (solid green line) and *dyz,zx* (solid pink line) orbitals of LaFeAsO$_{1-x}$H$_x$ with *x* = 0.08 (**k**), 0.21 (**l**), 0.36 (**m**) and 0.40 (**n**). The sum of the partial DOS of *dxy* and *dyz,zx* orbitals is also shown (dotted blue line).

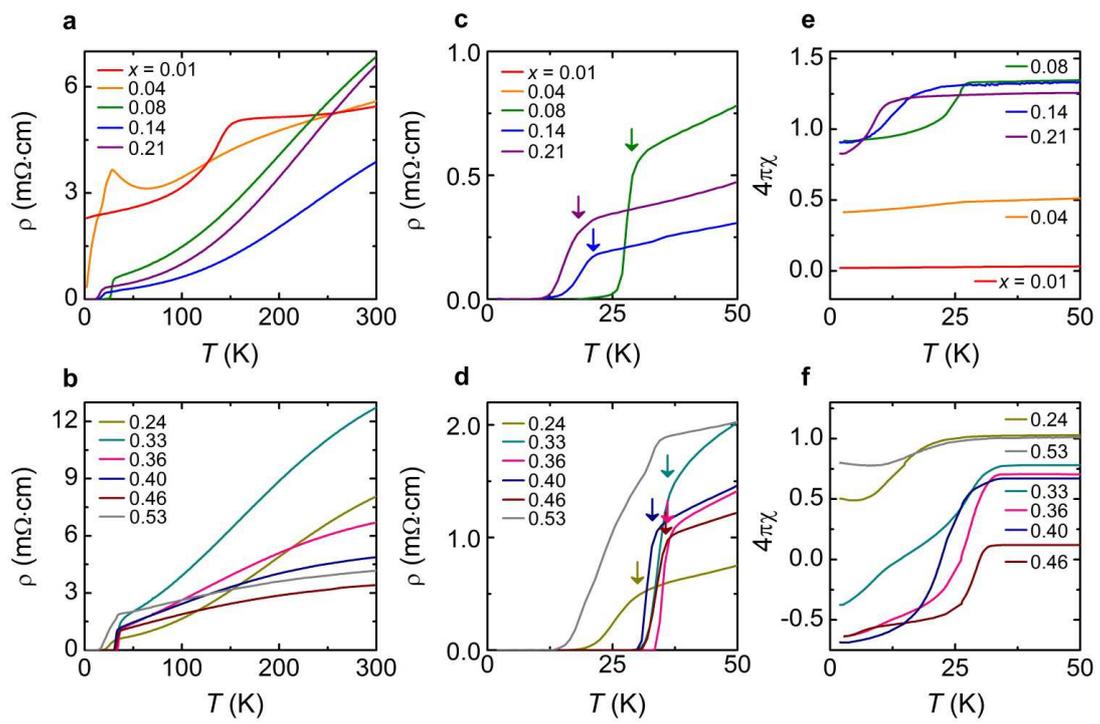

S. Iimura *et al.*

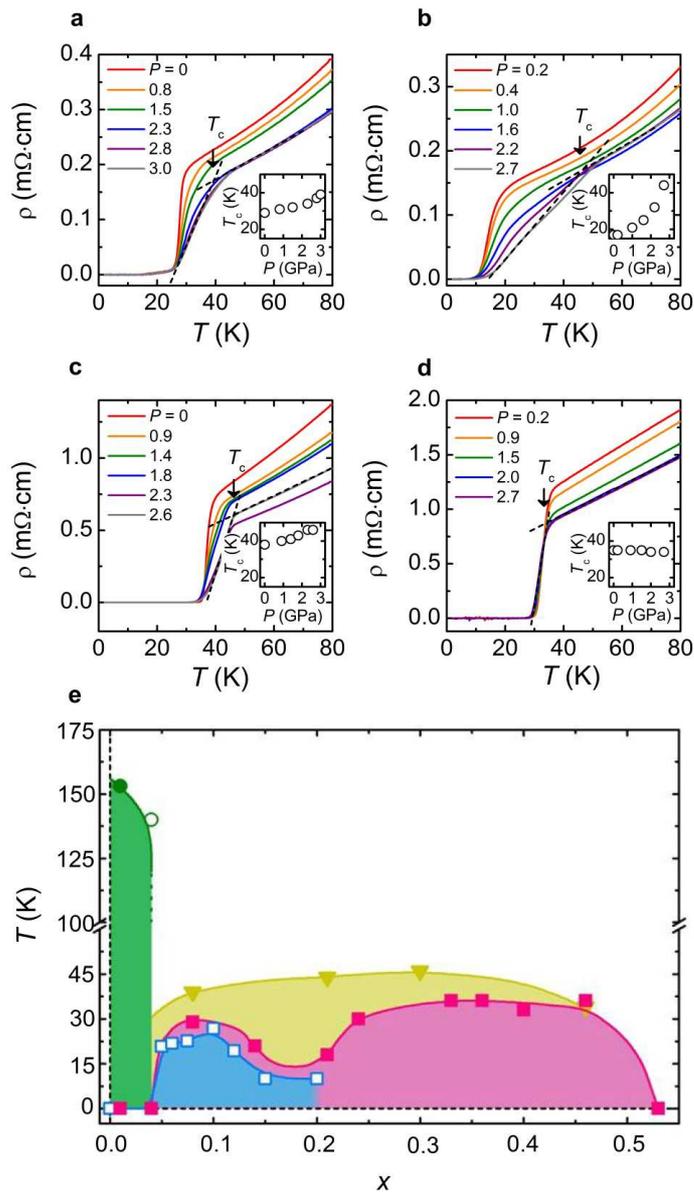

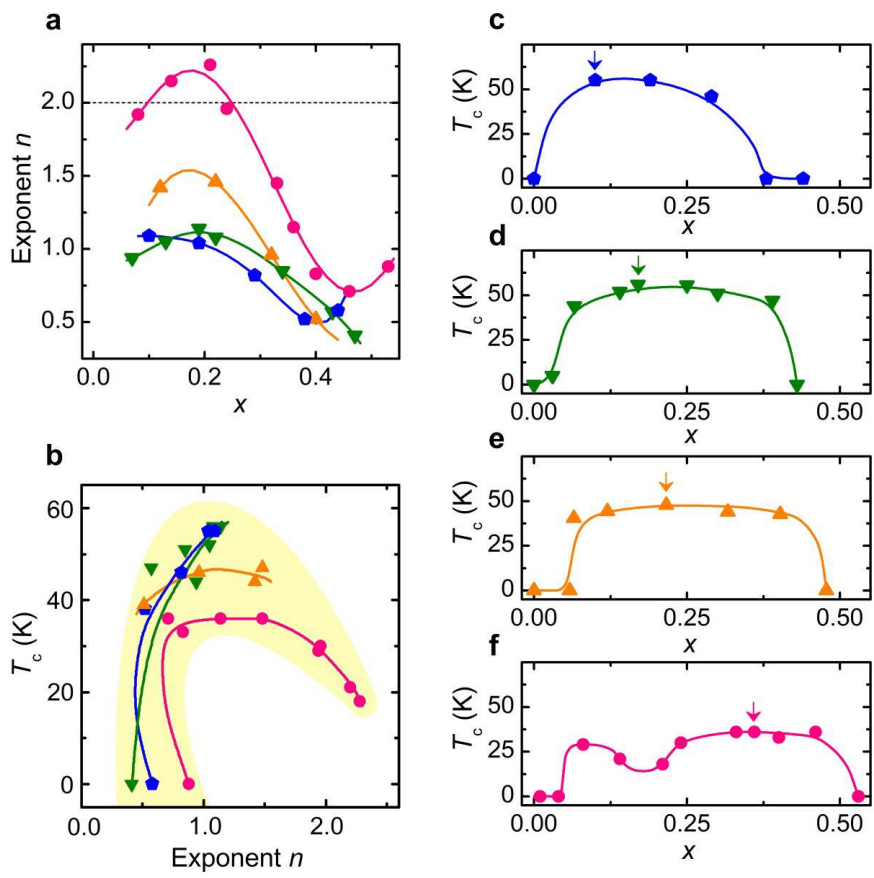

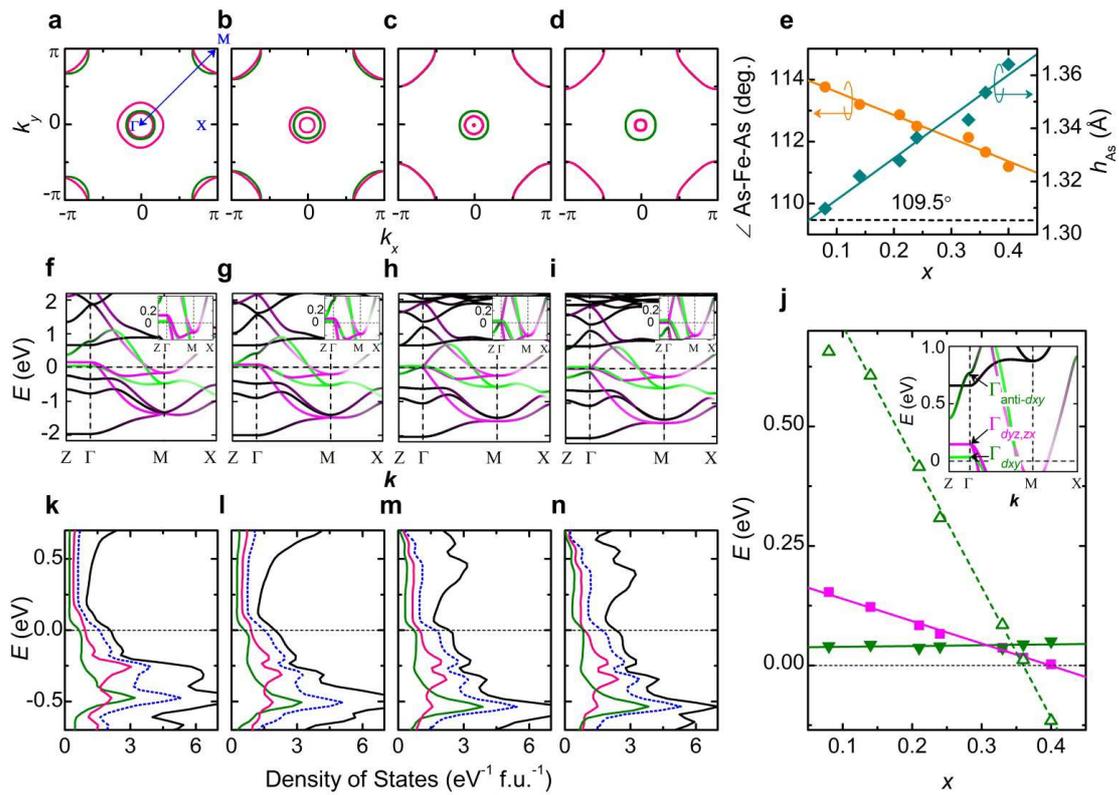

**Supplementary information**

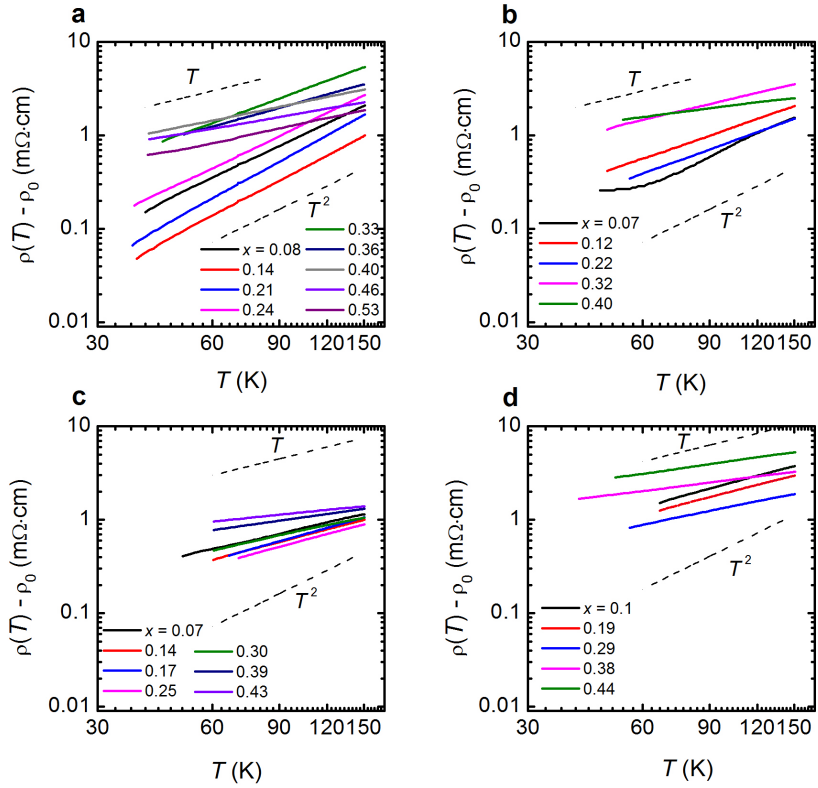

**Supplementary Figure S1 | Log [ρ(T) − ρ$_0$] vs. log T plots for $Ln$FeAsO$_{1-x}$H$_x$** The power-law fitting, $\rho(T) = \rho_0 + AT^n$ for $Ln$FeAsO$_{1-x}$H$_x$ ($Ln$ = (**a**) La, (**b**) Ce, (**c**) Sm and (**d**) Gd). The ρ$_0$ is the residual component of resistivity. The dashed line shows a visual guide showing $\rho(T) \sim T$ and $T^2$.

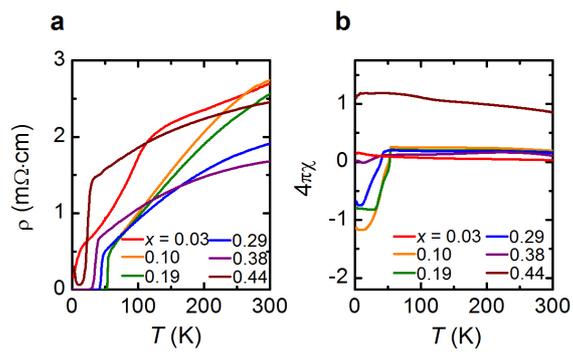

**Supplementary Figure S2 | Temperature dependence of electrical resistivity and volume magnetic susceptibility for GdFeAsO$_{1-x}$H$_x$** (a) Electrical resistivity as a function of temperature for GdFeAsO$_{1-x}$H$_x$. (b) Magnetic susceptibility data of GdFeAsO$_{1-x}$H$_x$ measured with a zero-field-cooling history and a field of 10 Oe.

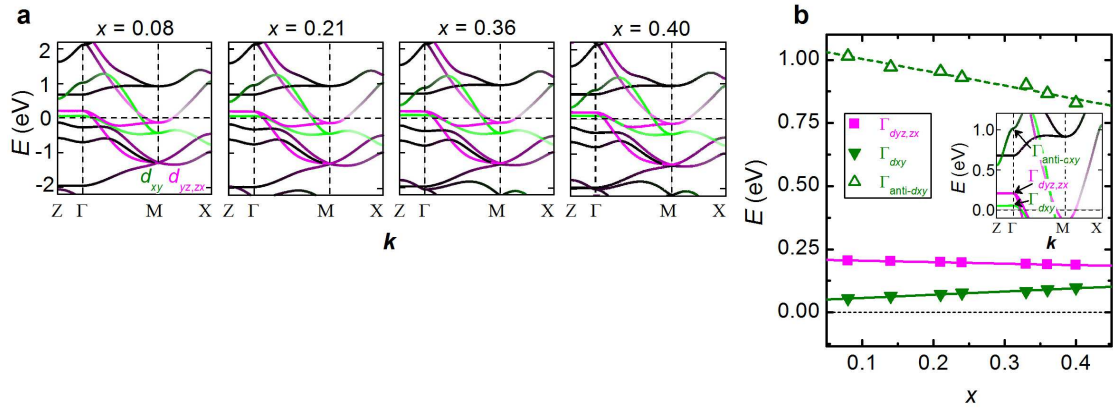

**Supplementary Figure S3 | Variation in band structure of LaFeAsO$_{1-x}$H$_x$ with only structural change** (a) Variation in band structure of LaFeAsO$_{1-x}$H$_x$ ($x$ = 0.08, 0.21, 0.36 and 0.40) with only structural change. The calculations were performed using the chemical formula, LaFeAsO, and the crystal structures of LaFeAsO$_{1-x}$H$_x$, This calculation procedure is similar to rigid band model, but we do not shift the $E_F$ with $x$. (b) Variation in energy level of relevant Fe 3$d$ bands at Γ point with $x$. The inset shows the band structure of LaFeAsO$_{0.92}$H$_{0.08}$. The solid and dashed lines are guides to the eye.

**SUPPLEMENTARY METHODS**

**Synthesis, structural and chemical analyses of GdFeAsO$_{1-x}$H$_x$.**

GdFeAsO$_{1-x}$H$_x$ was prepared by solid state reactions of Gd$_2$O$_3$, GdAs, GdH$_2$, FeAs and Fe$_2$As at 5 GPa and 1200C°. The phase purity and structural parameters at room temperature were determined by powder X-ray diffraction (XRD) measurements with MoKα1 radiation. Hydrogen content in the synthesized samples was evaluated by thermogravimetric mass spectroscopy, and the chemical composition with the exception of hydrogen was determined with a wavelength-dispersive-type electron-probe microanalyzer.